\documentclass[prl,showpacs,amssymb,floatfix,twocolumn]{revtex4}
\usepackage{amsmath}
\usepackage[usenames]{color} 
\usepackage{graphicx}
\usepackage{epsfig}
\usepackage{dcolumn}
\usepackage{bm}
\usepackage{ulem}
\usepackage{times}

\def\(({\left(}
\def\)){\right)}

\newcommand{\corrin}[1]{{\color{black}{#1}}}
\newcommand{\be}{\begin{equation}} \newcommand{\ee}{\end{equation}}
\newcommand{\bea}{\begin{eqnarray}} \newcommand{\eea}{\end{eqnarray}}
 \newcommand{\eps}{\varepsilon}
\newcommand{\s}{\sigma}

 \newcommand{\Tr}{{\rm Tr}}

\begin{document}

\title{Simple Glass Models and Their Quantum Annealing}

\author {Thomas J\"org$^1$, Florent Krzakala$^2$, Jorge Kurchan$^3$
  and Anthony C.\ Maggs$^{2}$} \affiliation{ $^1$ LPTMS, Universit\'{e} Paris-Sud, CNRS UMR
  8626, 91405 Orsay Cedex, France\\
  \corrin{$^2$ PCT, ESPCI ParisTech, 10 rue Vauquelin, CNRS UMR 7083 Gulliver, 75005 Paris, France} \\
  \corrin{$^3$ PMMH, ESPCI ParisTech, 10 rue Vauquelin, CNRS UMR 7636, 75005 Paris, France}}

\begin{abstract}
  We study first-order quantum phase transitions in mean-field spin
  glasses. We solve the quantum random energy model using elementary
  methods and show that at the transition the eigenstate suddenly
  projects onto the unperturbed ground state and \corrin{that} 
  the gap between the lowest states is exponentially small in the system 
  size. We argue that this \corrin{is} a generic feature of all 
  ``random first-order'' models, which
  includes benchmarks such as random satisfiability. We introduce a
  two-time instanton to calculate this gap in general, and discuss the
  consequences for quantum annealing.
\end{abstract}
  
  
\maketitle
\normalem

Solving hard combinatorial problems by temperature annealing is a
classic strategy in computer science \cite{Scott}. A major question is
whether annealing a \corrin{\em quantum mechanical} kinetic term \cite{QA,FG}
or a transverse magnetic field $\Gamma$ can be an efficient
strategy. Experimentally this question was studied in an Ising spin
glass (SG) \cite{Exp}; an archetype for difficult systems in physics. A
quantum first-order transition was observed at very low temperatures, as
had been previously found in model systems
\cite{Goldschmidt,QREM-finitep}. Here we address several open
questions: What is the underlying behavior of the wave functions at
the quantum spin glass transition?  Is quantum annealing efficient in
solving these difficult optimization problems?  We thus first solve a
simple quantum version of a spin glass model, the random energy model
(REM) \cite{PSPIN}. Despite its simplicity, it reproduces many
properties of mean-field field glasses \cite{PSPIN} and allows one to
model the behavior of a wide variety of phenomena such as the ``ideal''  glass
transition \cite{GLASS} and random heteropolymer folding
\cite{POLYMERS}. The REM also captures aspects of the phenomenology of
random satisfiability \cite{SAT} and is closely related to the random
code ensemble in coding theory \cite{BOOK}. All these problems belong
to the so-called ``random first-order'' (RFO), or ``one-step replica
symmetry breaking'' class.  To show that in all of these systems the minimal
spectral gap $\Delta$ between the ground and first excited states is
exponentially small in the size, we set up an instanton calculation
that allows one to compute the gap. The minimal spectral gap in turn
yields a lower bound $\tau \propto \Delta^{-2}$ \cite{QA} on the time
needed to find the ground state.

Quantum spin glasses have been investigated over the past $30$ years
\cite{BM} using an elaborate mathematical formalism combining the
replica~\cite{MPV} and the Suzuki-Trotter methods~\cite{Goldschmidt,Obuchi} 
in order to introduce disorder and quantum mechanics. The quantum 
transition has been found to be first order at low temperature for all RFO
models~\cite{Goldschmidt,QREM-finitep,Obuchi}. We show here first that the
quantum version of the random energy model (QREM) can be solved
analytically using only basic tools of perturbation theory, a derivation
whose simplicity provides a detailed understanding of \corrin{the} 
quantum glass transition. The minimal gap $\Delta$ is found to be
exponentially small in $N$. Next, we show that this result holds for all
RFO models, making quantum annealing an exponentially slow algorithm
in those cases.

\paragraph*{QREM Model.---}\hspace{-0.4cm} Consider $N$ Pauli spins $\s$ in a transverse 
field $\Gamma$ with the Hamiltonian:
\[
 {\cal{H}}(\{{\s\}}) = {\cal}E(\{{\s^z\}}) + \Gamma \sum_{i=1}^N
\s^x_{i} = {\cal{H}}_0 +\Gamma V\!,
\]
where ${\cal}E(\{{\s^z\}})$ is a function that takes $2^N$ different
values for the $2^N$ configurations of the $N$ spins. These values are
taken randomly from a Gaussian distribution of zero mean and variance
$N/2$, as in the REM \cite{PSPIN}. A concrete implementation is $
{\cal}E(\{{\s^z\}})= \lim_{\substack{ p \to \infty}}
\sum_{i_1,...,i_p} J_{i_1,...,i_p} \corrin{\s^z_{i_1} \cdots \s^z_{i_p}}$, where the
$ J_{i_1,...,i_p}$ are random Gaussian variables.  In the $\s^z$
representation ${\cal{H}}$ is a $2^N \times 2^N$ matrix whose diagonal
entries are the $2^N$ classical energies. The matrix elements
of ${\cal{H}}_{0}^{\alpha \alpha}=E_{\alpha}^{\corrin{\rm REM}}$ and
${\cal{H}}_{0}^{\alpha\neq \beta}= 0 $, while $V_{\alpha \beta} = 1$
if $\alpha$ and $\beta$ are two configurations that differ by a single
spin flip and zero otherwise. ${\cal{H}}$ is sparse and can be studied
numerically rather efficiently even for large
\corrin{matrix} sizes using Arnoldi and Ritz methods \cite{Ritz}

\paragraph*{Two easy limits.---}\hspace{-0.4cm} The model is trivially solved in the
limit $\Gamma \to 0$ and $\Gamma \to \infty$. For $\Gamma=0$, we
recover the classical REM with $N$ Ising spins and $2^N$
configurations, each corresponding to an energy level $E_{\alpha}$
\cite{PSPIN}: Call $n(E)$ the number of energy levels belonging to the
interval $(E,E+dE)$; its average over all realizations is easily
computed: $\overline{n(E)} = 2^N P(E) \propto
e^{N\((\ln\!2-E^2/N^2\))} = e^{Ns(E/N)}$, where $s(e)=\ln\!2-e^2$
(with $e=E/N$). There is therefore a critical energy density
$e_0=-\sqrt{\ln\!2}$ such that, if $e<e_0$, then with high probability
there are no configurations while if $e>e_0$ the entropy density is
finite. A transition between these two regimes arises at $\frac
1{T_c}=\frac{ds(e)}{de}\big|_{e_0}=2\sqrt{\ln\!2}$ and the
thermodynamic behavior follows: (\rm i) For $T<T_c$,
$f_{\rm REM}=-\sqrt{\ln\!2}$ and the system is frozen in its lowest energy
states. Only a finite number of levels (and only the ground state at
$T=0$) contribute to the partition sum. The energy gap between them is
finite. (\rm ii) For $T>T_c$, $f_{\rm REM}=-\frac 1{4T} - T \ln\!2$;
exponentially many configurations contribute \corrin{to} the partition sum.

In the opposite case of very large values of $\Gamma$, the REM contribution
to the energy can be neglected. In the $\s^x$ basis, we find $N$
independent classical spins in a field $\Gamma$; the entropy density
is just given by the log of a binomial \corrin{distribution} between 
$-\Gamma N$ and $+\Gamma N$ and the free-energy density is $f_{\rm para}=
-T\ln\!2-T\ln\!\((\cosh\!\Gamma/T\))$.

\begin{figure*}
  \hspace{-0.3cm}
  \includegraphics[width=0.73\columnwidth]{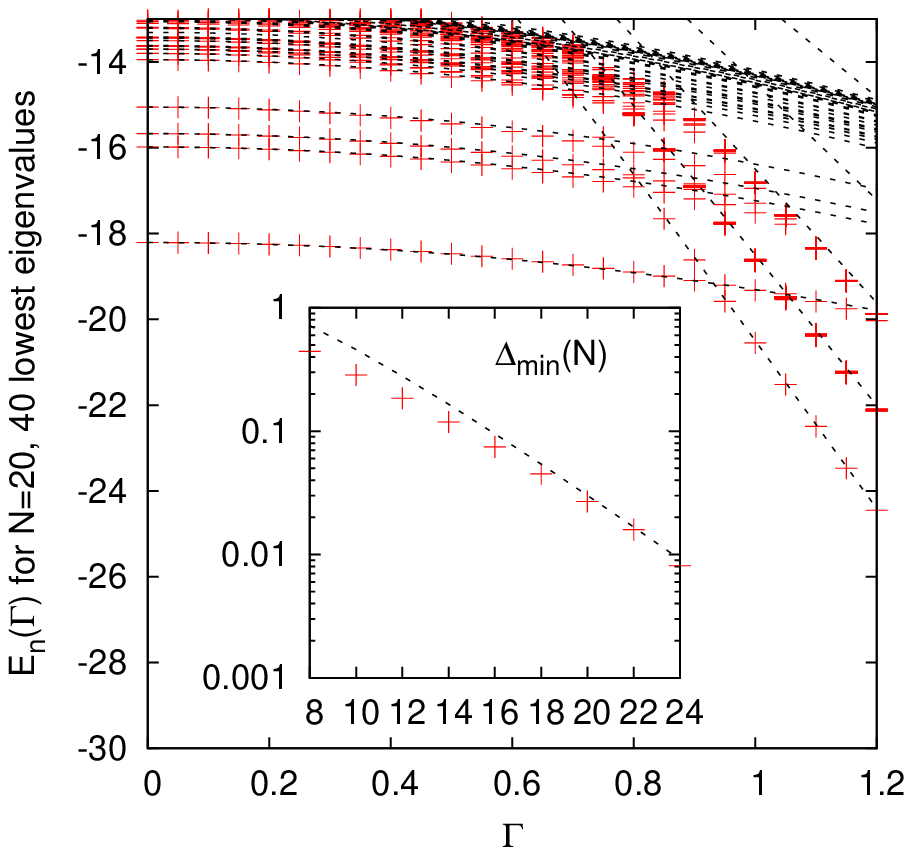}
  \hspace{-0.6cm}
  \includegraphics[width=0.73\columnwidth]{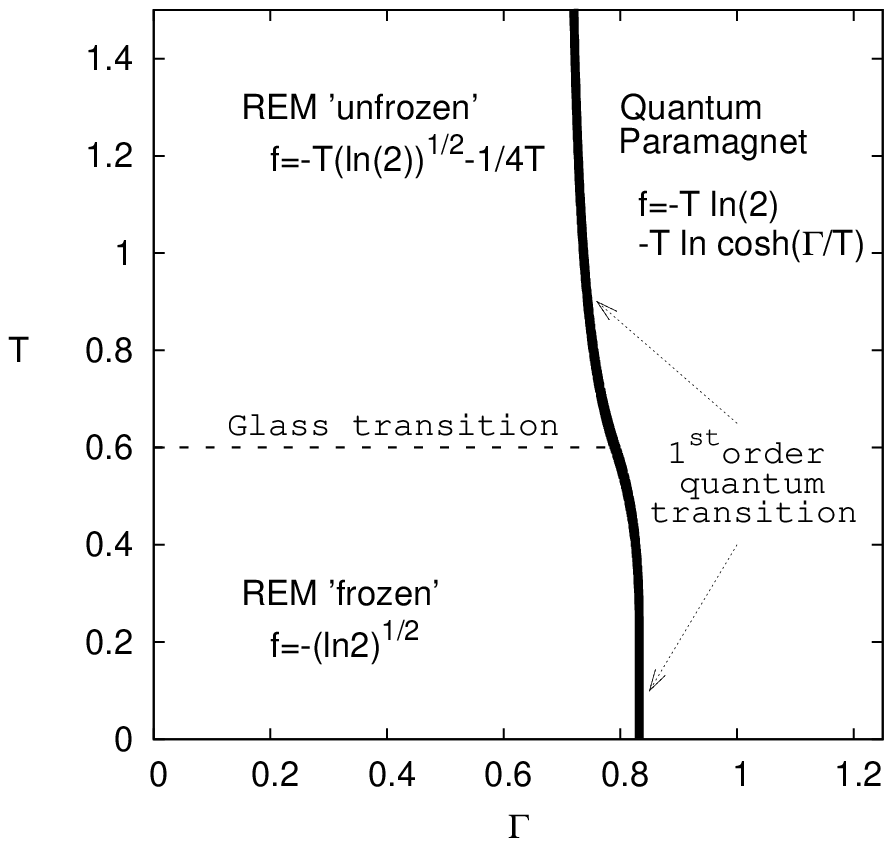}\hspace{-0.3cm}
  \includegraphics[width=0.70\columnwidth]{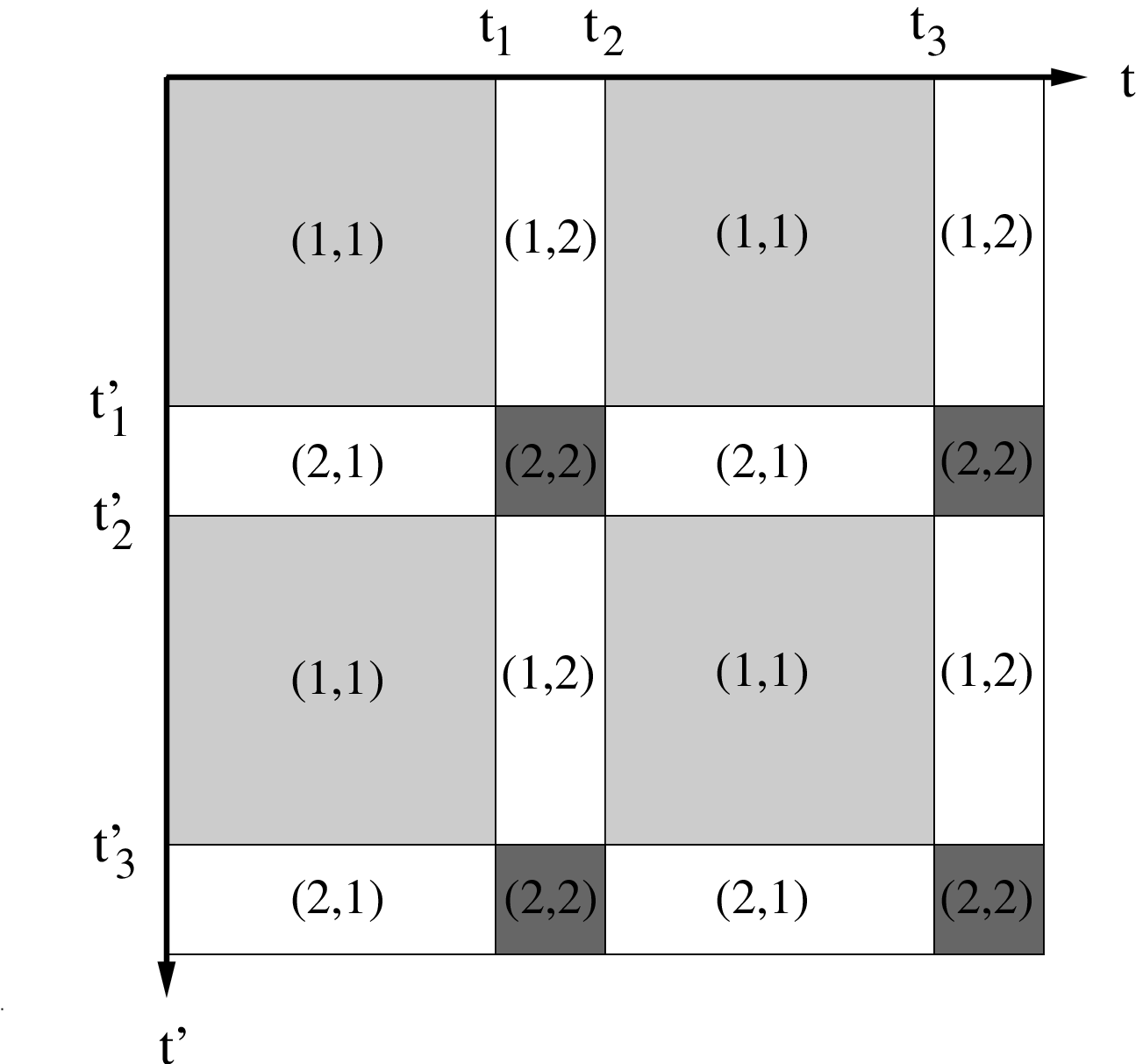}
  \vspace{-0.2cm}
\caption{(color online). Left: Evolution of lowest energy levels for 
  a single realization of the QREM with $N=20$ spins (dots) compared with
  analytical predictions (lines). 
 Inset: Evolution of the 
  ensemble averaged minimal gap at the transition. Center: Phase 
  diagram of the QREM in temperature $T$ and transverse 
  field $\Gamma$. At $T=0$ the quantum transition arises at 
  $\Gamma_c=\sqrt{\ln\!2}$ while the classical glass transition for 
  $\Gamma=0$ is at \corrin{$T_c=\sqrt{\ln\!2}/2$}. Right: A 
  multi-instanton configuration for the two-times overlap $q_{t,t'}$ 
  and the two-time Lagrange multipliers $\tilde{q}_{t,t'}$.  Far from
  the jump times, the functions $q_{t,t'}$ and $\tilde{q}_{t,t'}$ take
  the same form as those computed at those times for the glass phase
  [the regions (1,1)], for the quantum paramagnet (in the regions
  (2,2)), and are zero in the mixed regions (1,2)-(2,1). In the
  large $p$ limit the problem can be solved completely using the
  so-called ``static approximation'' \cite{Goldschmidt,Obuchi} within
  the (1,1) regions, and, in addition, the fact that ${\tilde
    q}^d_{t,t'}$ and $\tilde{q}_{t,t'}$ become either infinity or
  zero, with sharp interfaces.
  \label{levels}}
\end{figure*}

\paragraph*{Perturbation theory.---}\hspace{-0.4cm}What happens between these two
extreme cases? The perhaps surprising answer for the thermodynamics is 
{\it nothing}. At low value of $\Gamma$, the free-energy density is
that of the classical REM, while for larger values it jumps to the
quantum paramagnetic (QP) value $f_{\rm QP}$; a first-order transition
separates the two different behaviors at the value $\Gamma$ such that
$f_{\rm REM}=f_{\rm QP}$ (see center panel of Fig.~\ref{levels}). This 
can be easily understood using 
Rayleigh-Schr\"odinger 
perturbation theory~\cite{RS,foot1}.
Consider the set of eigenvalues $E_k$ and eigenvectors $|k \rangle$ of
the unperturbed REM, when $\Gamma=0$. The series for a given perturbed
eigenvalue $E_i(\Gamma)$ reads
\[ E_i(\Gamma)=E_i + \langle i|\!\sum_{n=0}^{\infty} \Gamma V \! \left[ \frac
{Q}{E_i-{\cal H}_{0}}(E_i-E_i(\Gamma)+\Gamma V) \right]^{\!n} \!\! |i \rangle,
\nonumber
\]
where the projector $Q=\sum_{\substack{k \neq i}} |k \rangle \langle
k|$ so that
\begin{align}
  &E_i(\Gamma)=E_i + \Gamma V_{ii} + \sum_{k \neq i} \frac{\Gamma^2
    V_{ik}V_{ki} }{E_i-E_k} + \cdots \,.
  \label{serie1}
\end{align}
Since $V_{ij} \neq 0$ if and only if $i$ and $j$ are two configurations
that differ by a single spin flip, odd order terms do not contribute
in Eq.~(\ref{serie1}) as one requires an even number of flips to come
back to the initial configuration in the sums. Noting that $\sum_{k
  \neq n} |V_{nk}|^2$ reduces to a sum over the $N$ levels connected
to $E_i$ by a single spin flip, one obtains, starting from an {\it
  extensive eigenvalue} [$E_i=O(N)$], that
\begin{equation} 
  \nonumber 
  \sum_{k \neq i} \frac{V_{ik}^2}{E_i-E_k} = 
  \frac{1}{E_i} \sum_{k=1}^N \left(1+\frac{E_k}{E_i}+\cdots \right) = 
  \frac{N}{E_i} + O\left(\frac{1}{N}\right)\!,  
\end{equation}
where we have used that the $E_k$ are random and typically of order
$\sqrt{N}$. Higher $n$th orders are computed in the same
spirit and are found to be $O(N^{n/2-1})$. Therefore, to all (finite)
orders, we have
\be E_i(\Gamma)= E_i + \frac{N\Gamma^2}{E_i} +
O\((\frac{1}{N}\)) \label{REM_perturb} \!. 
\ee
This analytic result compares well with \corrin{a} numerical evaluation 
of the eigenvalues (left panel of Fig.~\ref{levels}). Note that the energy
density of all extensive levels is independent of $\Gamma$ to leading
order in $N$ as are hence $s(e)$ and $f(T)$.

The expansion can also be performed using $\Gamma V$ as a starting
point and ${\cal H}_0$ as a perturbation. Consider the \corrin{ground
  state with} eigenvalue \corrin{$E_0(\Gamma)$ and the unperturbed
  ground state with $E_0^V(\Gamma)=-\Gamma N$}. In the base 
corresponding to the eigenvalues of $\Gamma V$, we find
\[ 
\corrin{E_0(\Gamma)} = \corrin{E_0^V(\Gamma)} + \langle \corrin{0} \mid
{\cal{H}}_0 \mid \corrin{0} \rangle + \sum_{k \neq \corrin{0}}
\frac{|\langle k \mid {\cal{H}}_0\mid \corrin{0} \rangle| ^2
}{\corrin{E_0^V(\Gamma)}-\corrin{E_k^V(\Gamma)}} + \cdots \,.
\]
The \corrin{first-order} term gives $\sum_{\corrin{\alpha=1}}^{2^N} E^{\rm REM}_\corrin{\alpha}
\corrin{|v_\alpha|^2}$. Since the energies of the REM are random and uncorrelated
with $v_\corrin{\alpha}$ this sums to $O(\sqrt{N}2^{-N/2})$. For the
\corrin{second-order} term, one finds
\corrin{
  \begin{align}
    \sum_{k \neq 0} \frac{|\langle k \mid {\cal{H}}_0\mid 0 \rangle|^2
    }{E_0^V(\Gamma)-E_k^V(\Gamma)} &= \frac{1}{E_0^V(\Gamma)} \sum_{k\neq 0}  \frac{|\langle k \mid {\cal{H}}_0\mid 0 \rangle|^2 }{1-E_k^V(\Gamma)/E_0^V(\Gamma)}
    \nonumber \\
    \approx \frac{1}{E_0^V(\Gamma)} \langle 0
    \mid {\cal{H}}_0^2 \mid 0 \rangle &= \frac{N}{2 E_0^V(\Gamma)} +
    o(1).
  \end{align} 
} 
Subsequent terms are treated similarly and give vanishing
corrections so that $\corrin{E_0}(\Gamma) = -N\Gamma - \frac
1{2\Gamma} + o(1)$. Again this \corrin{derivation} holds 
for other states with extensive energies \corrin{$E_i^V(\Gamma)$}, 
the only tricky point being the degeneracy of the
eigenvalues \cite{foot2}\corrin{, and for these excited eigenstates, the
  perturbation starting from the large $\Gamma$ phase yields $
  E_i(\Gamma) = E_i^V(\Gamma)-\frac 1{2\Gamma} + o(1)$.  Again, to leading 
  order in $N$, energy, entropy and free-energy densities are not 
  modified by the perturbation.
}

\paragraph*{The quantum transition.---}\hspace{-0.4cm}
This derivation sheds new light on the physics of the transition: The
wave function in the QP phase is delocalized over the classical
configurations in the $\s^z$ base. The first-order transition 
amounts to a sudden localization of the wave function into an
exponential number of classical states for $T>T_c$ and a finite number
of frozen states for $T<T_c$ (and to the ground state at $T=0$).

Focusing on $T=0$ and on the avoided level crossing near the
transition, we compute the gap $\Delta(N)$ as follows: Consider a
value of $\Gamma$ such that {\em for that sample} the SG
ground state and the quantum paramagnet are degenerate. We lift the
degeneracy by diagonalizing $\cal{H}$ in the corresponding
two-dimensional space
\be
{\cal{H}} |\phi \rangle = {\Large [}\corrin{E_0} |SG\rangle\langle SG| -
\Gamma N |QP\rangle\langle QP|  {\Large ]}|\phi \rangle = \lambda
|\phi \rangle.\nonumber 
\ee
\noindent
The gap is given by the difference of the eigenvalues, so that
\begin{equation}
  \Delta(N,\Gamma)^2= (N\Gamma-\corrin{E_0})^2- 
  4\left[ -\corrin{E_0} N\Gamma + \corrin{E_0}
    N\Gamma \langle SG|QP\rangle^2 \right] \nonumber
\end{equation}
and at the transition when $\Gamma=-\sqrt{\ln\!2}=E_0/N$, it yields
\be {\Delta}_{\rm min}(N)=2|\corrin{E_0}|2^{-N/2},
\label{GAP_}
\ee
where we have used the fact that $\langle SG|QP\rangle =
2^{-N/2}$. This agrees well with numerics, even for small values of
$N$ (see left panel of Fig.~\ref{levels}). Similar results are known
for number partitioning \cite{NP}.

\paragraph*{Generic case and Instanton.---}\hspace{-0.4cm}
A first-order quantum transition being a generic feature in all RFO
models, we expect these arguments to hold {\em qualitatively} in all
such models, so that the gap closes exponentially with $N$, much in
the same way that a {\em thermal} mean-field first-order transition implies
an exponential activation time and metastability. Indeed, quantum
annealing works by {\em tunneling} between quantum states, but in 
first-order transitions these states are usually ``far'' from each other. 
In order to {\em quantitatively} compute the gap, perturbation theory is
of no use in the generic case and one has to resort to instantonic 
computations\cite{Zinn}. We now discuss how this can be done in disordered 
systems using the replica method. To introduce the instanton, we use
the expansion of the evolution operator and, denoting 
$\eps = \langle QP|H|SG \rangle$, write
\be 
\Tr~e^{-\beta H} = \sum_{k~{\rm even}} \frac{1}{k!}  \int
\corrin{dt_1 \ldots dt_k} e^{-\left[t_{\rm tot}^{\rm SG} H_{\rm SG} + 
t_{\rm tot}^{\rm QP} H_{\rm QP}\right]}\; {\eps}^{k}\!,
\label{serie}
\ee
where the system jumps at $t_1,...,t_k$ between the states $|SG
\rangle$ and $|QP\rangle$, $t_{\rm tot}^{\rm SG}$ and $t_{\rm tot}^{\rm QP}$ 
is the total time spent in each. Following the standard strategy
\cite{Goldschmidt,Obuchi,QREM-finitep}, the trace is computed via the
Suzuki-Trotter and the replica trick. One obtains an effective
replicated free energy as a function of the overlaps $q^{\mu
  \nu}_{t,t'}$ between the replicas at two (imaginary) times and some
corresponding Lagrange multipliers $\tilde{q}^{\mu \nu}_{t,t'}$, for
which a particular ansatz must be proposed \cite{Goldschmidt,QREM-finitep}.  
Equation (\ref{serie}) tells us that if we find a solution that 
interpolates between $|SG \rangle$ and $|QP\rangle$ by jumping $k$ 
times $t_1,...,t_k$ and yields $\ln \Tr [e^{-\beta H}]\sim -t_{\rm tot}^{\rm SG} F_{\rm SG} -t_{\rm tot}^{\rm QP} F_{\rm QP}-k G$,
then by simple comparison $\ln \eps \sim G$ leads to $\Delta \sim
e^G$: An extensive value of $G$ implies an exponentially small gap and
the value of $G$ is thus proportional to the free-energy cost of an
interface in a two-time plane. For disordered systems, the computation 
can be performed by using a special two-time instanton ansatz as 
\corrin{shown} in the right panel of Fig~.\ref{levels}. We now 
refer to the presentation and notation of \cite{Obuchi}. We calculate 
the free energy per spin $f=F/N$ of the replicated systems in the 
$N\to\infty$ \corrin{limit} by the saddle point method. In the one-step replica 
symmetry ansatz, we divide replicas $\mu$ in $n/m$ sets of size $m$: 
we denote the parameters $q_{tt'}^{\mu \nu}$ as (\rm i) $q^d_{tt'}$ 
if $\mu=\nu$~\cite{foot7}, (\rm ii) $q_{tt'}$ if $\mu
\neq \nu$ but belong to the same block and zero otherwise.This
corresponds to the SG and the QP that have been widely studied 
\cite{Goldschmidt,Obuchi,QREM-finitep}:
\begin{align}
  -\beta f =&\int dt \; dt' \;\left\{
    -\frac{\beta^2 J^2}{4} (1-m)  q^p_{tt'} + \frac{(1-m)}{2}
    \tilde q_{tt'}   q_{tt'} \right. \nonumber \\ 
  &\left. + \frac{\beta^2 J^2}{4} [q^d_{tt'}]^p
    - \tilde q^d_{tt'}   q^d_{tt'} \right\}- \corrin{W_0}. \label{ooo} 
\end{align}
An expression for $\corrin{W_0}$ is given below. We consider a 
solution corresponding to the low-$\Gamma$ phase in the interval $(0,t_1)$,
$(t_2,t_3)$ that jumps to the high-$\Gamma$ \corrin{phase} in the 
intervals $(t_1,t_2)$, $(t_4,t_5)$, and so on.

As a proof of principle, let us rederive the large-$p$ case.
The saddle point equations imply that for large $p$ either 
$(q_{tt'},q^d_{tt'}, \tilde q_{tt'}, \tilde q^d_{tt'})=
(1,1,\infty,\infty)$ or $(q_{tt'},q^d_{tt'}, \tilde q_{tt'}, 
\tilde q^d_{tt'})=(<1,<1,0,0)$.  This implies that the form
of the instanton configuration of $\tilde q^d_{tt'}$ and $\tilde
q_{tt'}$ is the same as the one of $q_{tt'}$ and $q^d_{tt'}$ but 
with the values jumping from $0$ to $\infty$. In addition we 
make the {\em ``static approximation''} that assumes that 
inside each time interval the parameters $q^d$ and $\tilde q^d$ 
are constant. We conclude that we can write
\begin{equation}
  2\tilde q^d_{t't'} -\tilde q_{t't'} = r^d_t r^d_{t'}, \qquad 
  \qquad \tilde q_{t't'} = r_t r_{t'},
\end{equation}
where $r_t$ and $r^d_{t}$ are large in the time intervals when the
system is in the SG state, and drop to zero when it is not. (The
solutions in the literature correspond to a time-independent value of
$r$: large for the glass and small for the QP phase, respectively).
Because $ q^d_{t't'}, q_{t't'}$ are either zero or one, we have
\begin{align}
  &\int dt \; dt' \; q_{tt'} \tilde q_{tt'} \sim 
  \int dt \; dt' \; \tilde q_{tt'} = I^2 \nonumber \\ 
  2 &\int dt \; dt' \;\tilde q^d_{tt'}  q^d_{tt'} = 
  2 \int dt \; dt' \; \tilde q^d_{tt'} = I_d^2+ I^2
\end{align}
with the definitions $I \equiv \int dt \; r(t)$ and $ I_d \equiv \int
dt \; r^d(t)$. We further decouple the replicas in the single-spin
term in the usual way~\cite{Goldschmidt}:
\begin{align}
  \corrin{W_0}\!&=\!\ln \Tr \exp \left( -H_{{\rm eff}} \right) \nonumber \\ 
  \!&=\!-\frac{1}{m}\! \ln \!\left\{\!\int\!\!\!Dz_2\!\left[\!\int\!\!\!Dz_3 
      \Tr\!\left( {\cal{T}} 
        e^{\int\!dt' (A(t') \sigma^z + \beta \Gamma \sigma^x)} \right)
    \right]^{\!m} \!\right\}\!, \label{pp}
\end{align}
where $ {\cal{T}}$ denotes time order (a necessity here because of the
time dependence in the exponent), and $A(t) \equiv ( z_3 r^d_t+z_2
r_t)$. At low temperatures, the ``field'' in the $x$ direction 
$\beta \Gamma$ is strong, while the field in the $z$ direction 
$|A(t)|$ is either zero or $|A(t)|>>\beta \Gamma$. The single quantum 
spin then switches from being completely polarized along $|z\rangle$  
and along $|x\rangle$, in the periods in which $A \neq 0$ and $A=0$, 
respectively. The trace in (\ref{pp}) can then be calculated by 
switching the single-spin basis from $|x\rangle$ to $|z\rangle$.
Denoting $t^{\rm SG}=\Theta \beta$ the time when $q_t=q^d_t=1$, and
$t^{\rm QP}=(1-\Theta) \beta$ the rest, the action becomes
\begin{align}
  -\beta f =&\Theta^2 \;\left\{ -\frac{\beta^2 J^2}{4} (1-m)  
    +\frac{\beta^2 J^2}{4}   \right\} - \frac{1}{2} I_d^2 -
  \frac{m}{2} I^2  W_z \nonumber \\ 
  &+ (1-\Theta) \beta \Gamma +
  {\text{(number of jumps)}} \corrin{\times} \ln |\langle x|z \rangle|,
  \label{ppp}\nonumber
\end{align}
where \corrin{the} terms $|\langle x|z \rangle|$ come from \corrin{a} change of basis, and
\begin{equation}
  W_z=  
  -\frac{1}{m} \log \left\{ \int Dz_2 \left[ \int Dz_3 \;
      e^{|z_2  I +  z_3 I_d | } \right]^m \right\}\!\!.\nonumber
\end{equation}
This can be evaluated by the saddle point \cite{Goldschmidt,Obuchi}, a
short calculation yields $W_z \sim \frac{1}{2} I_d^2 + \frac{m}{2} I^2
+ \ln\!2 $.  Taking a further saddle point with respect to $m$ gives
$m=\frac{2\sqrt{2}}{\Theta \beta J}$ and thus
\begin{equation}
  -\beta f = 
  \Theta \sqrt{\ln\!2} \frac{\beta J}{2}   
  +  (1-\Theta) \beta \Gamma + k \ln |\langle x|z \rangle|.
  \label{ooo2}
\end{equation}
This is exactly the contribution to $\Tr[e^{-\beta H}]$ of the process
with $k$ jumps spending a fraction $\Theta $ in the glass state and
$(1-\Theta)$ in the paramagnetic state. We finally have $G=N \ln
|\langle x|z \rangle|= -N \ln(2)/2$ and we recover the result of
Eq.~(\ref{GAP_}).

In a generic problem with a first-order transition, one has to extremize 
the free energy (\ref{ooo}) and from there compute the gap as a free-energy 
cost of an interface that is generally nonzero.

\paragraph{Conclusion.---}\hspace{-0.4cm}
Starting from the quantum random energy model, we have discussed the 
quantum glass transition. The gap is exponentially small at the 
transition. We introduce a method that allows us to show that this
result holds for all models of the random first-order kind; presumably
including benchmark problems such as random satisfiability. Our 
results imply that quantum annealing is exponentially slow at finding 
the ground state of these random NP-hard 
(nondeterministic-polynomial-time-hard) problems. Although this seems 
to contradict recent numerical results \cite{Peter}, the problems 
considered there were {\em not} randomly chosen from a flat distribution 
and are therefore different from what has been considered in the present 
study and in the computer science literature of random constraint 
satisfaction problems.


\begin{thebibliography}{100}
\bibitem{foot2} Within each \corrin{degenerate} subspace 
  the perturbation ${\cal{H}}_0$ acts as a random orthogonal 
  matrix of variance $N2^{-N}$. This is tiny and weakly lifts the 
  degeneracy.
\bibitem{foot7} Except our
$(q^d_{tt'},\tilde{q}^d_{tt'})$, which are their
$(R_{tt'},\tilde{R}_{tt'})$.
\end{thebibliography}
\end{document}